\documentclass{mn2e}
\usepackage{epsf,times}
\newcommand{\etal}{{et al}\/.}
\begin{document}
\title[X-ray nuclei of radio sources]{The X-ray nuclei of
  intermediate-redshift radio sources}
\author[M.J.~Hardcastle \etal]{M.J.\ Hardcastle$^1$, D.A. Evans$^2$
  and J.H. Croston$^1$\\
$^1$ School of Physics,
  Astronomy and Mathematics, University of
Hertfordshire, College Lane, Hatfield, Hertfordshire AL10 9AB\\
$^2$ Harvard-Smithsonian Center for Astrophysics, 60 Garden Street, Cambridge, MA~02138, USA}
\maketitle
\begin{abstract}
We present a {\it Chandra} and {\it XMM-Newton} spectral analysis of
the nuclei of the radio galaxies and radio-loud quasars from the 3CRR
sample in the redshift range $0.1 < z < 0.5$. In the range of radio
luminosity sampled by these objects, mostly FRIIs, it has been clear
for some time that a population of radio galaxies (`low-excitation
radio galaxies') cannot easily participate in models that unify
narrow-line radio galaxies and broad-line objects. We show that
low-excitation and narrow-line radio galaxies have systematically
different nuclear X-ray properties: while narrow-line radio galaxies
universally show a heavily absorbed nuclear X-ray component, such a
heavily absorbed component is rarely found in sources classed as
low-excitation objects. Combining our data with the results of our
earlier work on the $z<0.1$ 3CRR sources, we discuss the implications
of this result for unified models, for the origins of mid-infrared
emission from radio sources, and for the nature of the apparent
FRI/FRII dichotomy in the X-ray. The lack of direct evidence for
accretion-related X-ray emission in FRII LERGs leads us to argue that
there is a strong possibility that some, or most, FRII LERGs accrete
in a radiatively inefficient mode. However, our results are also
consistent with a model in which the accretion mode is the same for
low- and high-excitation FRIIs, with the lower accretion luminosities
in FRII LERGs attributed instead to more efficient radio luminosity
production in those objects.

\end{abstract}

\section{Introduction}

\subsection{Unified models}

Radio galaxies and radio-loud quasars found in low-frequency-selected,
flux-limited samples such as 3CRR (Laing, Riley \& Longair 1983:
hereafter LRL) fall into a number of different observational classes
based on their radio and optical emission. Fanaroff \& Riley (1974)
showed that radio morphology can be related to total radio luminosity;
the well-known FRI/FRII division is at a 178-MHz
luminosity\footnote{Here, and throughout the paper, we use a
concordance cosmology with $H_0 = 70$ km s$^{-1}$ Mpc$^{-1}$,
$\Omega_{\rm m} = 0.3$ and $\Omega_\Lambda = 0.7$.} ($L_{178}$) of
approximately $5 \times 10^{24}$ W Hz$^{-1}$ sr$^{-1}$. Hine \&
Longair (1979) pointed out another difference between low-power and
high-power sources: at high radio luminosities, radio galaxies are
more likely to show strong nuclear optical narrow-line emission.
Following Laing \etal\ (1994) we refer to the weak-line objects, Hine
\& Longair's class B, as `low-excitation' radio galaxies (LERG), and
the strong-line, class A objects as `high-excitation' sources,
although there are some technical differences between the definitions
which we will discuss later. It is important to note that the
low-excitation/high-excitation division does not correspond to the
FRI/FRII division, although most FRIs are low-excitation objects and
most high-excitation objects are FRIIs. Finally, some high-excitation
objects show broad optical emission lines in addition to the narrow
lines already discussed. For reasons that are mostly historical, such
objects are called `broad-line radio galaxies' (BLRG) if the continuum
emission from the optical nucleus does not outshine the host galaxy,
and `radio-loud quasars' if it does. The lowest-luminosity object in
the 3CRR sample classed as a quasar has $L_{178} \sim 3 \times
10^{26}$ W Hz$^{-1}$ sr$^{-1}$. High-excitation sources without broad
emission lines are called narrow-line radio galaxies (NLRG).

Unified models for radio galaxies (e.g. Urry \& Padovani 1995) seek to
remove some of this observational complexity by proposing that some of
the different observational classes are intrinsically the same objects
viewed at different angles to the line of sight, with the differences
between them produced partly by the effects of beaming and partly (in
some cases) by the effects of an obscuring `torus' containing high
column densities of gas and dust. In this paper we
shall mostly be concerned with powerful objects, and it is now
generally accepted that unified models for powerful radio galaxies and
radio-loud quasars (e.g. Scheuer 1987; Barthel 1987, 1989) must be
true at some level: that is, at least some radio galaxies must be
quasars seen at an unfavourable angle, and so must host `hidden'
quasar nuclei.

Barthel (1989) examined 3CRR objects with $0.5 < z < 1.0$, and so with
$L_{\rm 178} \ga 7 \times 10^{26}$ W Hz$^{-1}$ sr$^{-1}$, and
concluded that it was possible that {\it all} objects classed as radio
galaxies could be quasars viewed at large angles to the line of sight.
At lower luminosities, this model has to be modified for two reasons.
Firstly, since there are no objects classed as quasars in the 3CRR
sample with $z<0.3$, but many luminous, FRII-type NLRG, the
narrow-line objects in 3CRR must be unified with some other class of
object. The unification counterparts of low-luminosity FRII NLRG are
almost certainly the BLRG (Laing \etal\ 1994; Hardcastle \etal\ 1998)
although some BLRG at higher luminosities may be objects seen at
viewing angles intermediate between quasars and NLRG. Secondly, while
there are almost no low-excitation FRII objects at high radio
luminosities, they make up a significant fraction of the population of
FRIIs in the two decades of radio luminosity between the FRI/FRII
boundary and the luminosity cutoff of Barthel (1989). Since the
narrow-line region is too large to be obscured, no change of
orientation can allow the low-excitation objects to appear as quasars
or BLRG, which always show high-excitation narrow lines. Instead, it
seems likely that these objects can appear at any angle to the line of
sight without changing their optical classification (except at very
small angles to the line of sight, where they may appear as FRII-type
BL Lac objects; Laing \etal\ 1994, Jackson \& Wall 1999, Hardcastle
\etal\ 2003). This picture is reinforced by studies of the optical
continuum from the nuclei of FRIIs (Chiaberge \etal\ 2002; Varano
\etal\ 2004) which show that LERGs have optical nuclei that lie on the
radio-optical correlation established for FRIs, implying that their
optical nuclei are largely unobscured and dominated by jet emission:
they argue that the lack of emission lines and of accretion-related
continuum (as seen in quasars and BLRG) might imply a low radiative
accretion efficiency for LERGs and FRIs, a point we return to later in
the paper.

Low-excitation radio galaxies have other properties that set them
apart from the population participating in the standard unified model:
their linear sizes and core prominences are differently distributed
(Laing \etal\ 1994, Hardcastle \etal\ 1998), their radio jets and
hotspots often have distinctive properties (Hardcastle \etal\ 1998)
and they often lie in denser environments as measured in the X-ray
(Hardcastle \& Worrall 1999) and optical (Hardcastle 2004). Until now,
though, it has not been possible to compare their hard X-ray emission
with that seen in NLRG, BLRG and quasars with sufficient spatial
resolution to make an unambiguous separation between nuclear and
extended emission. This comparison is the subject of the present
paper.

\subsection{X-ray emission from radio sources}

All active nuclei with conventional accretion discs are expected to be
intrinsically strong X-ray sources. In unified models, the torus is
required to obscure the optical and ultraviolet continuum and the
broad-line region of a quasar, and this material will also obscure any
X-ray emission from close to the accretion disc. The prediction of
unified models was therefore that objects unified with quasars (NLRG)
should show a component of heavily absorbed nuclear emission, and this
was borne out by early studies of individual objects (e.g. Ueno \etal\
1994) as well as more detailed studies of large samples with hard
X-ray instruments such as {\it ASCA} and {\it Beppo-SAX} (e.g.
Sambruna, Eracleous and Mushotzky 1999; Grandi, Malaguti \& Fiocchi 2006). The work of
Sambruna \etal\ also hinted that LERG might have lower nuclear X-ray
luminosities than NLRG and broad-line objects, although their sample
was heterogeneous. In addition, however, work in the soft X-ray,
largely with {\it ROSAT}, showed that radio-loud AGN had a component
of nuclear X-ray emission that was {\it not} heavily obscured even in
NLRG, and was well correlated with the nuclear radio (`core') emission
(e.g. Worrall \& Birkinshaw 1994; Worrall \etal\ 1994; Edge \&
R\"ottgering 1995; Canosa \etal\ 1999; Trussoni \etal\ 1999) implying
a relationship with the parsec-scale jet. Although this relationship
was best studied in nearby, low-power FRI radio sources, where it was
possible to use radio, optical and X-ray data to investigate the
relationship between FRI sources and BL Lac objects (e.g. Capetti
\etal\ 2000; Hardcastle \& Worrall 2000), the radio-X-ray correlation
persisted in FRII radio galaxies (e.g. Hardcastle \& Worrall 1999),
implying that these more powerful sources had similar jet-related
nuclear X-ray properties.

Recently studies with {\it Chandra} and {\it XMM-Newton} have
confirmed this picture. Using (FRII) 3CRR sources in the redshift
range $0.5 < z < 1.0$, where the standard unified model works well,
Belsole, Worrall \& Hardcastle (2006) reproduce with {\it Chandra} and
{\it XMM} the radio core/soft X-ray correlation found by Hardcastle \&
Worrall (1999) with {\it ROSAT}, while also finding that a large
fraction of the NLRG in their sample show an additional heavily
absorbed component, as the unified model would predict. Low-redshift
($z<0.1$) 3C and 3CRR sources, largely FRIs, have been studied by
several groups (Donato \etal\ 2004; Evans \etal\ 2006 (hereafter E06);
Balmaverde \etal\ 2006): all find that essentially all sources show a
soft component of nuclear emission whose luminosity correlates with
that of the radio core, as seen with {\it ROSAT}. E06 found that the
narrow-line FRIIs in their sample (which included all 3CRR sources
with $z<0.1$) showed an additional heavily absorbed nuclear component,
which was accompanied by Fe K$\alpha$ line emission, and which is
most likely to originate near the accretion disc. The properties of
the narrow-line FRIIs were thus consistent with the expectation from
unified models.

However, the various analyses show that {\it no} FRI radio galaxy in
the 3CRR sample shows any evidence for this type of heavily absorbed
nuclear component. Indeed, only one FRI radio galaxy, Cen A, is known
to have an energetically dominant component with an absorbing column
greater than a few $\times 10^{22}$ cm$^{-2}$ (e.g. Evans \etal\
2004). This result has several possible interpretations. If FRI nuclei
generally show no absorbed nuclear emission, does this mean that the
torus is absent, or that a standard Shakura-Sunyaev accretion disc is
absent, or simply that both are present but that the accretion-related
X-rays are too weak to be seen? The fact that the FRI and FRII sources
lie on the same radio core/X-ray correlation, when the unabsorbed
components of the FRIIs are considered, strongly argues that we cannot
simply infer that the torus is absent in FRIs while the accretion disc
is still present: if this were the case, we would expect the FRIs to
have an additional component of unabsorbed X-ray emission, and to lie
above the correlation, which is not observed. E06 put constraints on
the luminosities of any undetected heavily absorbed components in the
FRIs, and showed that they are systematically lower than those of
FRIIs, even for sources of comparable radio power, which might argue
in favour of a true difference between the accretion luminosities (and
hence possibly accretion modes) in FRIs and FRIIs, as previously
proposed (e.g.\ Ghisellini \& Celotti 2001). We return to this point
later in the current paper.

E06 raised the question of whether there exists a population of FRIIs
which, like the FRIs, have jet-dominated X-ray emission. An obvious
candidate population is the low-excitation radio galaxies, but in the
E06 sample there was only one LERG (3C\,388) and, although its
properties were consistent with having a weak or absent heavily
absorbed component, the nuclear X-ray spectrum was of low quality.
Thus both the poorly studied nuclear X-ray properties of LERGs, and
the possible similarity between LERGs and the FRI sources studied by
E06 and others, motivated us to extend the E06 sample to higher redshifts and
obtain more X-ray spectra of both LERG and NLRG radio galaxies. The
results of this investigation are the subject of the present paper.

\subsection{This paper}

In this paper we begin by collating the available {\it Chandra} and
{\it XMM-Newton} data on the 3CRR sources in the redshift range $0.1 <
z < 0.5$, thus filling the redshift and luminosity gap between the
samples of E06 and Belsole \etal\ (2006). We then combine the new data
with those of E06 and explore the consequences for the position of
LERG in unified schemes, and the nature of the accretion flow in LERG
and FRIs.

\section{Data and analysis}

\subsection{Sample}
\label{class}
There are 50 objects in the 3CRR sample with $0.1 < z < 0.5$. 16 of
these have been observed with {\it Chandra} and 7 with {\it
XMM-Newton}; 3 have been observed with both, giving a total of 20
sources with useful X-ray data. The X-ray observations cover 6/12 of
the objects classed as LERG, 7/26 of the NLRG, 3/7 of the BLRG and 5/5
of the quasars in the sample (see below for more information on
classifications), and so are heavily biased away from NLRG; {\it
Chandra}, in particular, has only observed 2 NLRG in this redshift
range. We took an early decision to omit 3C\,48 from our analysis,
since it is a compact steep-spectrum source with peculiar morphology
and not obviously unified with either FRIs or FRIIs, and since the
short {\it Chandra} observation has been discussed in detail elsewhere
(Worrall \etal\ 2004). The remaining quasars in the sample are all
lobe-dominated with FRII radio morphology. We include X-ray data from
two additional $z<0.1$ FRII radio galaxies, 3C\,192 and 3C\,285, which
were not available to E06, and, since
E06 did not determine an explicit upper limit on absorbed emission in
3C\,388, we re-analyse that as well. The 22 sources included in our
analysis are listed in Table \ref{sources}. Redshifts and
emission-line classifications are taken from the 3CRR catalogue (LRL)
with updates from the online
version\footnote{http://3crr.dyndns.org/}.

It is worthwhile to comment explicitly on the emission-line
classifications we use in this paper. These are derived from LRL, and
subsequent work on the sample, notably some of the work discussed by
Laing \etal\ (1994); thus, at least to some extent, they reflect a
{\it qualitative} assessment of the nature of the emission-line
spectrum, comparable to the classes of Hine \& Longair (1979). We take
`weak-lined' or `absorption-only' sources in LRL to be LERG in the
absence of high-quality spectral data such as those of Laing \etal\
(1994), and `strong-lined' sources to be NLRG. Jackson \& Rawlings
(1997) give classifications for some, but not all, sources in the 3CRR
sample from data available in the literature, using the quantitative
classification of Laing \etal\ (1994), i.e. that equivalent width of
the [OIII] line is less than 10~\AA\ or that the line radio
[OII]/[OIII] $>1$. The Jackson \& Rawlings classifications of FRIIs
agree in the vast majority of cases with the qualitative 3CRR
classifications that we use for sources in our sample and that of E06,
though they do not provide complete coverage of FRIIs with $z<0.5$.
The one disagreement is over 3C\,438, which is classed by LRL as
absorption-line only (following Smith \& Spinrad 1980) but which
Jackson \& Rawlings class as a high-excitation NLRG. The evidence for
detected [OIII] in 3C\,438 (Rawlings \etal\ 1989) is very weak,
however, and its flux is certainly low, and so for consistency we
retain 3C\,438 in the LERG class. The online table\footnote{
http://www-astro.physics.ox.ac.uk/$\sim$cjw/3crr/3crr.html} collated
by Willott \etal\ (1999) disagrees with both Jackson \& Rawlings and
LRL in classifying 3C\,388 as a high-excitation object, it classifies
3C\,295 (unclassified by Jackson \& Rawlings) as low-excitation, and
it puts two of our sample objects (3C\,79 and 3C\,223) into the class
of weak quasars, which is equivalent to our BLRG class, again
disagreeing with Jackson \& Rawlings' classification. Their
classification of 3C\,79 and 3C\,223 is based entirely on nuclear
infrared emission -- they do not suggest that broad emission lines are
directly detected -- and hence we feel justified in following LRL and
Jackson \& Rawlings for these objects. Their classification of 3C\,388
refers to Rawlings \etal\ (1989), who in turn refer to Saunders \etal\
(1989), who report an estimated [OII]/[OIII] $>1$, so 3C\,388 is a
LERG, as correctly reported by Jackson \& Rawlings. The Willott \etal\
classification of 3C\,295, however, is based on the high-quality
spectrum of Lawrence \etal\ (1996), which does show [OII]/[OIII] $>1$.
Varano \etal\ (2004) argue that 3C\,295's optical properties are more
consistent with those of NLRG, and suggest that the source may have
been misclassified. For consistency with the classification of the
other sources we retain 3C\,295 in the NLRG class, but we will
explicitly note situations where a reclassification would make a
difference to our conclusions.

\begin{table*}
\caption{Sources in the sample, radio and optical information, and
  X-ray observational data. The columns headed log$_{10}(L_{\rm 178})$
  and log$_{10}(L_{\rm 5})$ give the log to base 10 of the 178-MHz
  total luminosity and 5-GHz core luminosity, respectively, in units
  of W Hz$^{-1}$ sr$^{-1}$. Radio data are taken from the online 3CRR
  catalogue. Where three livetimes are quoted, it indicates that {\it
  XMM-Newton} data were used: in this case the livetimes are for the
  MOS1, MOS2 and pn instruments, in that order.}

\label{sources}
\begin{tabular}{lrrrrrrrrll}
\hline
Source&$z$&178-MHz&$\alpha$&log$_{10}$($L_{178}$)&5-GHz&log$_{10}$($L_{5}$)&Emission&Galactic&Observation&Livetime\\
&&flux density&&(W Hz$^{-1}$&core flux&(W Hz$^{-1}$&line type&$N_{\rm H}$&ID&(s)\\
&&(Jy)&&sr$^{-1}$)&(mJy)&sr$^{-1}$)&& ($\times 10^{20}$ cm$^{-2}$)\\
\hline
3C\,28    & 0.1952  & 17.8 & 1.06 & 26.19 & $<$0.2 & $<$21.16 & LERG & 5.14 &3233 &49720\\
3C\,47    & 0.425   & 28.8 & 0.98 & 27.17 & 73.6 & 24.43 & Q    & 5.71 &2129 &44527\\
3C\,79    & 0.2559  & 33.2 & 0.92 & 26.72 & 10 & 23.10 & NLRG &10.09 &0201230201 &16651, 16894, 12255\\
3C\,109   & 0.3056  & 23.5 & 0.85 & 26.73 & 263 & 24.68 & BLRG &15.61 &4005 &45713\\
3C\,123   & 0.2177  & 206.0 & 0.70 & 27.33 & 100 & 23.96 & LERG &43.00 &829 &47016\\
3C\,173.1 & 0.292   & 16.8 & 0.88 & 26.55 & 7.4 & 23.09 & LERG & 5.25 &3053 &23999\\
3C\,192   & 0.0598  & 23.0 & 0.79 & 25.19 & 8 & 21.71 & NLRG & 5.06 &0203280201 &8559, 8582, 6335\\
3C\,200   & 0.458   & 12.3 & 0.84 & 26.86 & 35.1 & 24.17 & LERG & 3.69 &838 &14660\\
3C\,215   & 0.411   & 12.4 & 1.06 & 26.78 & 16.4 & 23.75 & Q    & 3.68 &3054 &33803\\
3C\,219   & 0.1744  & 44.9 & 0.81 & 26.47 & 51 & 23.47 & BLRG &10.38 &827 &18756\\
3C\,223   & 0.1368  & 16.0 & 0.74 & 25.79 & 9 & 22.50 & NLRG & 1.20 &0021740101 &25776, 27382, 18376\\
3C\,249.1 & 0.311   & 11.7 & 0.81 & 26.44 & 71 & 24.13 & Q    & 2.89 &0153210101 &15379, 17399, 12796\\
3C\,284   & 0.2394  & 12.3 & 0.95 & 26.22 & 3.2 & 22.55 & NLRG & 1.00 &0021740201 &43063, 43099, 35806\\
3C\,285   & 0.0794  & 12.3 & 0.95 & 25.18 & 6 & 21.84 & NLRG & 1.37 &6911 &39624\\
3C\,295   & 0.4614  & 91.0 & 0.63 & 27.70 & 3 & 23.11 & NLRG & 1.38 &2254 &90936\\
3C\,303   & 0.141   & 12.2 & 0.76 & 25.70 & 150 & 23.75 & BLRG & 1.72 &1623 &14951\\
3C\,346   & 0.162   & 11.9 & 0.52 & 25.80 & 220 & 24.04 & NLRG & 5.47 &3129 &44413\\
3C\,351   & 0.371   & 14.9 & 0.73 & 26.71 & 6.5 & 23.25 & Q    & 2.03 &2128 &50920\\
3C\,388   & 0.0908  & 26.8 & 0.70 & 25.63 & 62 & 22.97 & LERG & 5.81 &5295 &30711\\
3C\,401   & 0.201   & 22.8 & 0.71 & 26.30 & 32 & 23.39 & LERG & 7.42 &3083, 4370 &47518\\
3C\,436   & 0.2145  & 19.4 & 0.86 & 26.31 & 19 & 23.22 & NLRG & 8.20 &0201230101 &26427, 27833, 22190\\
3C\,438   & 0.290   & 48.7 & 0.88 & 27.00 & 7.1 & 23.07 & LERG &17.22 &3967 &47272\\

\hline
\end{tabular}
\end{table*}

\subsection{{\it Chandra} analysis}

The archive data were uniformly reprocessed from the level 1 events
file with {\sc ciao} 3.3 and CALDB 3.2. The latest gain files were
applied and the 0.5-pixel randomization removed using standard
techniques detailed in the {\sc ciao} on-line
documentation\footnote{http://asc.harvard.edu/ciao/}. No time
filtering was carried out apart from applying the standard good time
intervals, since the object was to study point sources, which should
not be significantly affected by high background. All but one of our
targets had a clear detected point source coincident with the nucleus
of the host galaxy and the radio core. (The exception, 3C\,28, is
discussed in Appendix A.) For sources that
were not piled up, spectra of the X-ray nuclei were extracted in
2.5-pixel circles (1 pixel is 0.492 arcsec) centred on the visible
point-like component, with background taken between 2.5 and 4.0 pixels
using the {\it specextract} script, which also builds the appropriate
response files. The small extraction region and local background
subtraction ensures that extended thermal emission from the host
galaxy, group or cluster is largely removed before fitting. Spectra
were grouped, typically to 20 counts per bin after background
subtraction.

{\it Chandra} observations of several of the sources (3C\,47, 3C\,109,
3C\,215, 3C\,219, and 3C\,303) are affected by photon
pileup. We demonstrated this by calculating the 0.5--7 keV count rates
from source-centred circles of radius 1.23 arcsec (2.5 pixels).
Typical values are in the range 0.35--0.4 counts per frame, for which
a pileup fraction 15--20 per cent is predicted by the {\sc PIMMS} software.
We estimated the spatial extent of the pileup by producing images
consisting of grade~7 events, which are largely produced by photon
pileup. Inspection of the images showed the pileup to be concentrated
within the central 1 arcsec. In order to sample spectra free from
pileup, we extracted spectra from the wings of the PSF using
source-centred annuli of inner radius 1 arcsec (2.0 pixels) and outer radius 2
arcsec (4.1 pixels), with background sampled from surrounding annuli of inner
radius 2 arcsec and outer radius 3 arcsec. We corrected the point-like
ancillary response files (ARFs) for the energy-dependent missing flux
using the {\sc arfcorr}
software\footnote{http://agyo.rikkyo.ac.jp/$\sim$tsujimot/arfcorr.html}
that calculates the encircled energy fraction in an annular extraction
region of a model PSF created using {\sc chart} and {\sc
marx}\footnote{http://cxc.harvard.edu/chart/}.

\subsection{{\it XMM} analysis}

The archive {\it XMM-Newton} data were reduced using the {\it
XMM-Newton} Scientific Analysis Software (SAS) package. The data were
filtered for good time intervals using a count-level threshold
determined by examining a histogram of the off-source count rate in
the energy ranges 10-12 keV (MOS) and 12-14 keV (pn). GTI filtering
resulted in a mean reduction in exposure time of $\sim 15$ per cent,
with the worst case (3C249.1) having a reduction of 30 per cent. The
data were then filtered using the flag bitmasks 0x766a0600 for MOS and
0xfa000c for pn, which are equivalent to the standard flagset
\#XMMEA\_EM/EP but include out of field-of-view events and exclude bad
columns and rows. They were also filtered for the standard selections
of patterns $\le 12$ (MOS) and $\le 4$ (pn).

Spectral analysis was performed using scripts based on the {\sc sas}
{\it evselect} tool to extract spectra from all three events lists. We
used on-axis response files and ancillary response files generated
using the {\sc sas} tasks {\it rmfgen} and {\it arfgen} (with
encircled energy corrections included). All {\it XMM} targets had a
detected compact central X-ray component. Core spectra were extracted
from circles of radius 35 arcsec, with background spectra extracted
from off-source regions on the same CCD chip. As for the {\it Chandra}
data, spectra were grouped to a minimum of 20 
background-subtracted counts per bin prior to spectral fitting.

\subsection{Spectral fitting}

Spectral fitting was carried out using {\sc xspec} 11 in the energy
range 0.4-7.0 keV (for {\it Chandra} data) or 0.3-8.0 keV ({\it
XMM-Newton}). For consistency, we followed the same fitting procedure
for every source. First, a single power law with fixed Galactic
absorption was fitted to the data (see Table \ref{sources} for the
Galactic column densities used, which are largely derived from the
{\sc colden} software). In one or two cases there was a clear
requirement for an excess absorption at soft energies, and in this
case a free absorbing column, assumed to be at the redshift of the
source, was added to the model. For those sources where a second
component of X-ray emission was clearly required (seen in large
residuals and poor $\chi^2$ values) we then added a second power law
with a free, but initially large, absorbing column at the redshift of
the source. This gave rise to good fits in almost all the sources
where the $\chi^2$ was initially poor, though occasionally it was
necessary to fix the index of the unabsorbed power-law to $\Gamma =
2.0$ to get good constraints on its normalization. (The choice of
photon index for the frozen power law clearly affects the resultant
normalization, but our tests show that the effect is seen at the 10
per cent level at most for plausible values of $\Gamma$.) Errors on the
parameters were determined using $\Delta\chi^2 = 2.7$ (90 per cent
confidence for one interesting parameter) for consistency with E06. If
a single power law provided a good fit to the data, we added a heavily
absorbed power law with fixed $\Gamma = 1.7$ and an absorbing column
$N_{\rm H} = 10^{23}$ cm$^{-2}$ at the redshift of the source and
re-fitted. The fixed parameters of this component, particularly the
absorption column, are consistent with what is found in sources with
detected heavily absorbed components, and were chosen to agree with
the choices of E06; we comment later on the consequences of varying
them. If the 90 per cent uncertainty on the normalization of this new
component was consistent with zero, we treated the upper bound as an
upper limit on a heavily absorbed component. If, on the other hand,
the fit was improved with a non-zero normalization for this component,
we allowed the absorbing column and, if well-constrained, the photon
index of the second power law to vary, and treated the resulting model
as a detection of a second component. This in practice only happened
for two quasars, 3C\,47 and 3C\,249.1 (see Appendix~A), and clearly in
these cases the data would be equally well fitted with a broken
power-law model, but we retain the results of the fits for
consistency. Additional components, such as a Gaussian around rest
energies of 6.4 keV or thermal emission, were added if required by the
residuals, as discussed in Appendix A.

We emphasise that the resulting fits are not unique. In many cases the
soft component of the spectrum, which we fit with an unabsorbed power
law, is only present at the level of one or two spectral bins. In
these cases an unabsorbed thermal component with suitably chosen
temperature and abundance would clearly give just as good a fit. Even
where there are many counts in the unabsorbed component, as in some of
our {\it XMM} datasets, a thermal model with a high temperature can
often provide just as good a fit. Our motivations for excluding
thermal models in these cases are based on physical plausibility
rather than directly on the X-ray spectra. Similarly, we cannot
exclude a partial covering origin for the unabsorbed component,
although of course the covering fraction in this case would have to
vary widely from source to source. Similarly, a scattering model
cannot be ruled out. Our motivation for using the model we adopt is
based on the existing evidence for a jet-related origin for the soft
component; we discuss the evidence for the validity of our approach in
the next section.

\section{Results}

\begin{table*}
\caption{Results of spectral fitting. For each source the fits, or
  upper limits, for an absorbed and unabsorbed power law are shown.
  The model described is the best-fitting model (PL = power law,
  ABS(PL) = absorbed power law, TH = thermal component, GAU =
  Gaussian). Components of the fit other than power-law components are
  discussed in Appendix A. Numerical values marked with a dagger were
  frozen in the fit (either to derive upper limits or because the data
  were not good enough to constrain them. 1-keV flux densities and
  luminosities are the unabsorbed values in all cases.}
\label{results}
\begin{tabular}{lllllllll}
\hline
Source&Net counts&Model&$\chi^2/$d.o.f.&Component&1-keV flux&Photon
&log$_{10}$&$N_{\rm H}$\\
&&&&&(nJy)&index&luminosity&($\times 10^{22}$\\
&&&&&&&(ergs s$^{-1}$)&cm$^{-2}$)\\
\hline
3C\,28    & $<$19 & PL &-- &PL&$<$0.5&--&$<$41.36 &--\\
&&&&ABS(PL)&$<$3.0&--&$<$42.27 &10.0\dag\\
3C\,47    & 1135 $\pm$ 39& PL+ABS(PL) &52.2/44 &PL&363$_{-33}^{+28}$ & 1.91$_{-0.21}^{+0.24}$ & 45.01 &--\\
&&&&ABS(PL)&308$_{-146}^{+146}$ & 1.70\dag & 45.05 &10.7$_{-5.6}^{+17.6}$\\
3C\,79    & 202 $\pm$ 17, 208 $\pm$ 17, & PL+ABS(PL) &48.7/44 &PL&9.5$_{-1.6}^{+1.6}$ & 2.87$_{-0.30}^{+0.31}$ & 42.42 &--\\
&478 $\pm$ 25&&&ABS(PL)&145$_{-102}^{+401}$ & 1.77$_{-0.61}^{+0.69}$ & 44.18 &24.9$_{-8.3}^{+11.0}$\\
3C\,109   & 2611 $\pm$ 57& ABS(PL) &68.2/89 &ABS(PL)&896$_{-109}^{+129}$ & 1.62$_{-0.11}^{+0.12}$ & 45.23 &0.5$_{-0.1}^{+0.1}$\\
&&&&ABS(PL)&$<$233.3&--&$<$44.60 &10.0\dag\\
3C\,123   & 610 $\pm$ 26& PL + ABS(PL) &21.7/20 &PL&1.9$_{-1.6}^{+1.5}$ & 2.00\dag & 42.00 &--\\
&&&&ABS(PL)&31.8$_{-12.8}^{+8.3}$ & 1.41$_{-0.33}^{+0.49}$ & 43.58 &3.1$_{-1.0}^{+1.5}$\\
3C\,173.1 & 26 $\pm$ 5& PL &-- &PL&0.8$_{-0.3}^{+0.3}$ & 2.00\dag & 41.55 &--\\
&&&&ABS(PL)&$<$8.6&--&$<$43.13 &10.0\dag\\
3C\,192   & 93 $\pm$ 11, 66 $\pm$ 10, & TH+POW+ABS(POW) &21.5/12 &PL&5.9$_{-2.1}^{+1.9}$ & 1.88$_{-0.55}^{+0.19}$ & 41.37 &--\\
&211 $\pm$ 17&&&ABS(PL)&159$_{-55}^{+87}$ & 1.70\dag & 42.93 &51.6$_{-16.6}^{+32.8}$\\
3C\,200   & 189 $\pm$ 14& PL &7.3/7 &PL&9.1$_{-1.2}^{+1.2}$ & 1.72$_{-0.25}^{+0.27}$ & 43.58 &--\\
&&&&ABS(PL)&$<$14.0&--&$<$43.78 &10.0\dag\\
3C\,215   & 740 $\pm$ 30& PL &34.6/28 &PL&231$_{-17}^{+17}$ & 1.80$_{-0.11}^{+0.11}$ & 44.84 &--\\
&&&&ABS(PL)&$<$86.0&--&$<$44.46 &10.0\dag\\
3C\,219   & 304 $\pm$ 19& PL &5.6/11 &PL&259$_{-31}^{+31}$ & 1.90$_{-0.18}^{+0.19}$ & 43.99 &--\\
&&&&ABS(PL)&$<$207.3&--&$<$44.02 &10.0\dag\\
3C\,223   & 404 $\pm$ 25, 416 $\pm$ 26, & PL+ABS(PL+GAU) &84.7/76 &PL&43.5$_{-4.2}^{+3.5}$ & 1.62$_{-0.21}^{+0.19}$ & 43.16 &--\\
&984 $\pm$ 42&&&ABS(PL)&34$_{-15}^{+101}$ & 0.71$_{-0.61}^{+0.77}$ & 43.67 &5.7$_{-3.6}^{+7.8}$\\
3C\,249.1 & 9066 $\pm$ 97, 10372 $\pm$ 103, & PL+ABS(PL+GAU) &971.6/970 &PL&618$_{-184}^{+72}$ & 2.29$_{-0.14}^{+0.30}$ & 44.72 &--\\
&27819 $\pm$ 170&&&ABS(PL)&167$_{-85}^{+170}$ & 1.25$_{-0.26}^{+0.25}$ & 44.74 &0.4$_{-0.3}^{+0.4}$\\
3C\,284   & 140 $\pm$ 16, 159 $\pm$ 17, & PL+ABS(PL) &51.9/34 &PL&4.0$_{-0.3}^{+0.3}$ & 2.37$_{-0.15}^{+0.15}$ & 42.22 &--\\
&580 $\pm$ 31&&&ABS(PL)&98$_{-91}^{+340}$ & 1.70\dag & 43.98 &161.7$_{-47.2}^{+379.1}$\\
3C\,285   & 396 $\pm$ 20& PL+ABS(PL+GAU) &6.4/13 &PL&0.4$_{-0.2}^{+0.2}$ & 2.00\dag & 40.40 &--\\
&&&&ABS(PL)&227$_{-49}^{+60}$ & 1.70\dag & 43.33 &32.1$_{-4.6}^{+5.5}$\\
3C\,295   & 474 $\pm$ 29& PL+ABS(PL+GAU) &16.8/15 &PL&1.0$_{-0.5}^{+0.5}$ & 2.00\dag & 42.50 &--\\
&&&&ABS(PL)&82$_{-9}^{+172}$ & 1.85$_{-0.20}^{+0.25}$ & 44.48 &41.0$_{-8.4}^{+7.8}$\\
3C\,303   & 245 $\pm$ 18& PL &6.3/9 &PL&223$_{-31}^{+31}$ & 1.60$_{-0.21}^{+0.21}$ & 43.91 &--\\
&&&&ABS(PL)&$<$227.8&--&$<$43.86 &10.0\dag\\
3C\,346   & 3045 $\pm$ 57& PL &99.4/91 &PL&58.5$_{-2.0}^{+2.0}$ & 1.70$_{-0.05}^{+0.05}$ & 43.40 &--\\
&&&&ABS(PL)&$<$6.4&--&$<$42.44 &10.0\dag\\
3C\,351   & 8493 $\pm$ 93& PL+ABS(PL) &235.9/201 &PL&7.2$_{-2.4}^{+3.3}$ & 5.13$_{-0.48}^{+0.50}$ & 41.92 &--\\
&&&&ABS(PL)&167.1$_{-11.8}^{+7.1}$ & 1.42$_{-0.10}^{+0.11}$ & 44.80 &0.8$_{-0.1}^{+0.1}$\\
3C\,388   & 183 $\pm$ 16& PL &4.1/5 &PL&9.3$_{-1.7}^{+1.7}$ & 2.22$_{-0.29}^{+0.30}$ & 41.74 &--\\
&&&&ABS(PL)&$<$8.3&--&$<$42.01 &10.0\dag\\
3C\,401   & 406 $\pm$ 29& PL &16.9/16 &PL&7.5$_{-0.9}^{+0.8}$ & 1.64$_{-0.22}^{+0.21}$ & 42.74 &--\\
&&&&ABS(PL)&$<$16.7&--&$<$43.05 &10.0\dag\\
3C\,436   & 192 $\pm$ 17, 183 $\pm$ 17, & PL+ABS(POW+GAU) &32.1/35 &PL&7.9$_{-0.7}^{+0.6}$ & 2.03$_{-0.19}^{+0.24}$ & 42.59 &--\\
&564 $\pm$ 29&&&ABS(PL)&44$_{-15}^{+24}$ & 1.70\dag & 43.53 &36.2$_{-13.2}^{+19.9}$\\
3C\,438   & 78 $\pm$ 13& PL &1.4/2 &PL&1.3$_{-0.6}^{+0.6}$ & 1.07$_{-0.59}^{+0.56}$ & 42.67 &--\\
&&&&ABS(PL)&$<$9.2&--&$<$43.14 &10.0\dag\\

\hline
\end{tabular}
\end{table*}

Results of the spectral fits for each source are given in Table
\ref{results}. Individual sources are discussed, and references to
previous work given, in Appendix A. Most of the fits resulting from
the procedure we describe above have good reduced $\chi^2$ values,
suggesting that the models give adequate representations of the data. 
For the few sources where the fit is less good, we see no evidence in
the residuals for any {\it systematic} devation from the models, such
as might arise from a missing component in the fits that was present in
a number of different sources. We therefore feel able to adopt the
qualitative and quantitative results of all the fits in the discussion
that follows.

Table \ref{results} immediately shows an interesting difference
between the FRII LERG and NLRG. Whereas all (8/8) of the NLRG in
our new sample show evidence for a heavily absorbed, luminous component in
hard X-rays, consistent with what was observed for the low-redshift
sample by E06, all but one (6/7) of the LERG show no such component,
and the exception (3C\,123) may not be a LERG at all (Section
\ref{123}). One LERG (3C\,28) is undetected, one (3C\,173.1) is barely
detected, and four others (3C\,200, 3C\,388, 3C\,401, 3C\,438) are well fitted
with power-law models. If 3C\,295, which does show evidence for a
heavily absorbed nuclear component, were to be reclassified as a LERG
(Section \ref{class}) then the statistics would be 7/7 NLRG and 6/8
LERG behaving as described above, so the trend is still noticeable.

\begin{figure}
\epsfxsize 8.4cm
\epsfbox{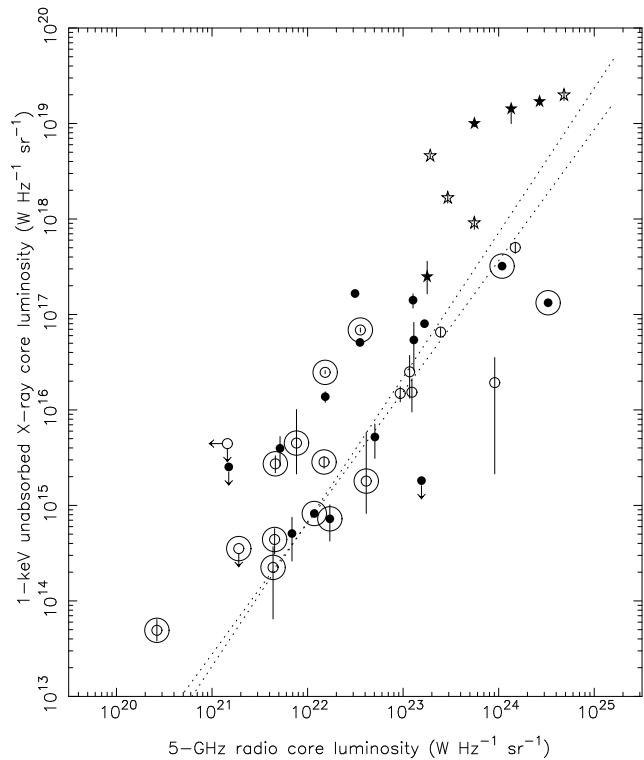}
\caption{X-ray luminosity of the unabsorbed component for the combined
  $z<0.5$ sample as a function of 5-GHz radio core luminosity (Table
  \ref{sources}). Open circles are LERG, filled circles NLRG, open
  stars BLRG, and filled stars quasars. Large surrounding circles
  indicate that a source is an FRI. Error bars are the 90 per cent
  confidence limits described in the text. Where error bars are not
  visible they are smaller than symbols. Dotted lines show the 90 per
  cent confidence range of regression lines fitted through the NLRGs
  and LERGs using the maximum-likelihood techniques described by
  Belsole \etal\ (2006). BLRGs and quasars lie above this
  correlation.}
\label{radio-xray}
\end{figure}

\begin{figure}
\epsfxsize 8.4cm
\epsfbox{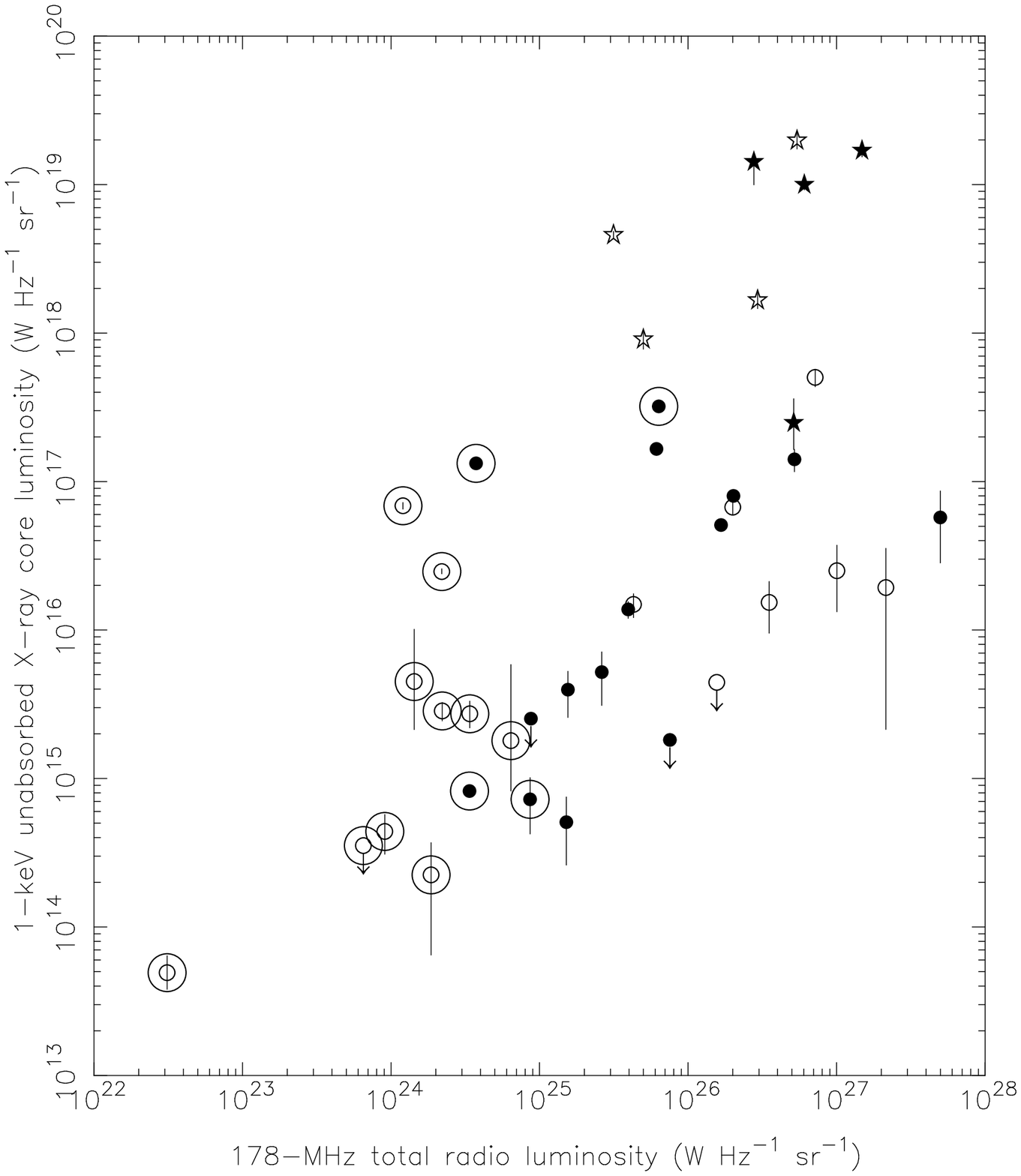}
\caption{X-ray luminosity of the unabsorbed component for the combined
  $z<0.5$ sample as a function of 178-MHz total radio luminosity (Table
  \ref{sources}). Symbols as in Fig.\ \ref{radio-xray}.}
\label{178-xray}
\end{figure}

For quantitative analysis, we combine our sample with the 3CRR sources
discussed in E06 to give a total sample size of 40 X-ray observed
objects with $z<0.5$ (out of 86 in total); this allows comparisons to
be made between the LERGs in our sources and the FRIs in E06. We begin
by plotting the 1-keV luminosity of the low-absorption power-law
component against the 5-GHz luminosity of the radio core, derived on
the assumption of a flat radio spectral index (Fig. \ref{radio-xray}). This
illustrates that the correlation between the soft X-ray emission and
the radio core, already discussed by Hardcastle \& Worrall (1999), E06
and Belsole \etal\ (2006), is, unsurprisingly, still present in the
$z<0.5$ sample. As has been found previously (Hardcastle \& Worrall
1999) the broad-line objects in the sample lie above the trendline
established by NLRG and LERG (which is attributed to the presence of
unabsorbed accretion-related emission in the spectra of the BLRG and
quasars). However, there is no systematic difference in the behaviour
of FRIs and FRIIs, or NLRG and LERG, although there is a good deal of
scatter about the line. Fig. \ref{178-xray} shows the equivalent plot
using the total 178-MHz luminosity of the source, illustrating that
the correlation is worse when the X-ray luminosities are plotted
against an unbeamed quantity; though both are formally well correlated
(at $>99.9$ per cent significance on a Kendall's T test), the
correlation with radio core luminosity is better than with total
luminosity, and a partial correlation analysis (partial Kendall's $T$,
neglecting the few upper limits) with redshift as the third variable
shows a significant positive partial correlation between the
quantities of Fig. \ref{radio-xray}, but not between those plotted in
Fig.\ \ref{178-xray}. The arguments that the correlation between X-ray
emission and core radio flux requires that the soft X-ray component of
the FRIs and NLRG/LERG FRIIs originate in the nuclear jet have been
discussed in detail by Hardcastle \& Worrall (1999, 2000) and E06,
among others, and we will not repeat them here. The key point of
relevance to the present paper is that the LERG occupy the same region
of parameter space in Fig. \ref{radio-xray}, and also in other
quantities such as the photon index of the fitted power law, as the
FRIs and NLRG FRIIs. This strongly suggests that the bulk of the
unabsorbed X-ray emission in LERGs (and in the other sources)
originates in a jet rather than in an unobscured or partially obscured
AGN. Although other models (such as a scattering model) could be
devised which might reproduce the X-ray/radio core correlation for a
particular class of source, if we accept that the FRI X-ray emission
originates in a jet (Section 1.2) then any alternative model requires
a coincidence to explain the fact that the {\it normalizations} of the
correlations are indistinguishable in all types of source.

\begin{figure}
\epsfxsize 8.4cm
\epsfbox{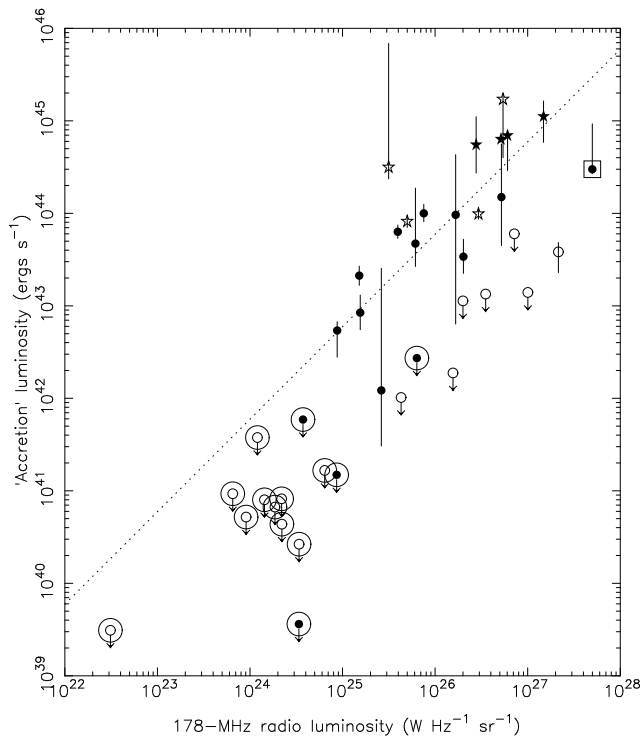}
\caption{X-ray luminosity of the accretion-related component for the combined
  $z<0.5$ sample as a function of 178-MHz total radio luminosity (Table
  \ref{sources}). Symbols as in Fig.\ \ref{radio-xray}. The dotted line
  is a line of slope unity whose normalization is determined by the
  median ratio between the two luminosities for the detected NLRG. A
  box marks 3C\,295, whose status as a NLRG is in doubt.}
\label{lum-lum}
\end{figure}

We now wish to consider the component of X-ray emission that may be
related to accretion. The offset of the lobe-dominated quasars and
broad-line objects in our sample that are fitted with a single
power-law model from the radio core-soft X-ray luminosity relationship
(Fig. \ref{radio-xray}) suggests that in these sources an
accretion-related component is dominant; thus, here, the total 2-10
keV luminosity is a reasonable estimate of the accretion-related 2-10
keV luminosity (though ideally we would subtract a contribution from a
jet-related component). For the NLRG and the few quasars where a
two-component model is a good fit, we can take the accretion-related
luminosity to be the unobscured luminosity of the absorbed power-law
component: this is supported by the presence of Fe K$\alpha$ emission
in a few of the NLRG. In all cases this accretion luminosity is
comparable to or larger than (for the NLRG, typically an order of
magnitude larger than) the luminosity of the unabsorbed or weakly
absorbed component.

Estimating the
accretion-related luminosity of the LERGs is harder simply because
there is no {\it a priori} reason to suppose that an obscuring torus
exists in these objects (a point we return to later in the paper) so
that we do not know whether, or to what extent, any accretion-related
luminosity in the LERGs is obscured. However, the 2-10 keV
absorption-corrected luminosity we estimate as an upper limit on an
obscured component with $N_{\rm H} = 10^{23}$ cm$^{-2}$ is in almost
all cases substantially larger than the luminosity of the detected,
unobscured component (Table \ref{results}). The luminosities could be
even larger if the obscuring columns are very much larger in the
non-detected objects than in the detected ones: columns of $10^{24}$
cm$^{-2}$ and higher can increase the `hidden' X-ray luminosities by
an order of magnitude or more. As E06 point out, such high column
densities and correspondingly high nuclear luminosities are ruled out
by infrared data in FRIs (e.g. M\"uller \etal\ 2004). The available
data are much sparser for LERGs, but we comment on some infrared
constraints in Section \ref{ir}. In the meantime, we adopt the limits
based on a column density of $10^{23}$ cm$^{-2}$, and refer to these,
together with the detections for the NLRG, as
the `accretion-related luminosity' in what follows; the reader should
bear in mind that the upper limits depend on a particular choice of
absorbing column, and in particular that they are highly conservative
if there is no absorption at all.

A plot of the 2-10 keV accretion-related luminosity against 178-MHz radio
luminosity then shows a clear difference between LERGs and NLRGs (Fig.
\ref{lum-lum}). The upper limits on the accretion-related components
in the LERGs, given our assumed absorbing column of $10^{23}$ cm$^{-2}$, lie
systematically below the detected NLRG at all radio powers. In this
respect they resemble the FRIs discussed by E06. Both FRIs and FRII LERGs
tend to lie below a line of slope unity that passes through the NLRGs.
If we interpret the absorbed component in NLRG as directly related to
accretion, then this implies that the LERGs and FRIs have lower
accretion rates for a given radio luminosity. Only an extreme
absorbing column for the LERGs ($>10^{24}$ cm$^{-1}$) could bring them
back into agreement with the NLRGs. We discuss the
implications of this result in the following section.

\section{Discussion}

\subsection{The nature of LERGs and the FRI/FRII dichotomy}

If we neglect for the time being the possibility of an extreme absorbing
column in the LERGs, the fact that there is a class of FRII objects
(which for convenience we simply refer to as the LERGs in what
follows) that have lower accretion power than the NLRGs for a given
radio luminosity can be interpreted in two ways. If 178-MHz radio
power can be taken to reflect jet power, $L_{\rm jet}$, then LERGs and
NLRGs matched in radio luminosity produce the same $L_{\rm jet}$ for a
lower accretion luminosity, $L_{\rm acc}$; this could be used, as in
model 3 of E06, to argue for a different mode of accretion in the
LERGs. If, on the other hand, the observed $L_{\rm acc}$ is an
indicator of the true $L_{\rm jet}$, then LERGs (and FRIs) manage to
produce a greater amount of radio emission for a given $L_{\rm jet}$ ;
at a minimum, the LERGs would have to produce about ten times the
radio power for a given jet power. The latter model would predict that
accretion-related emission might be found at a level predictable from
the jet power in LERGs, while in the former model accretion-related
emission might not be detected at all.

\begin{figure}
\epsfxsize 8.4cm
\epsfbox{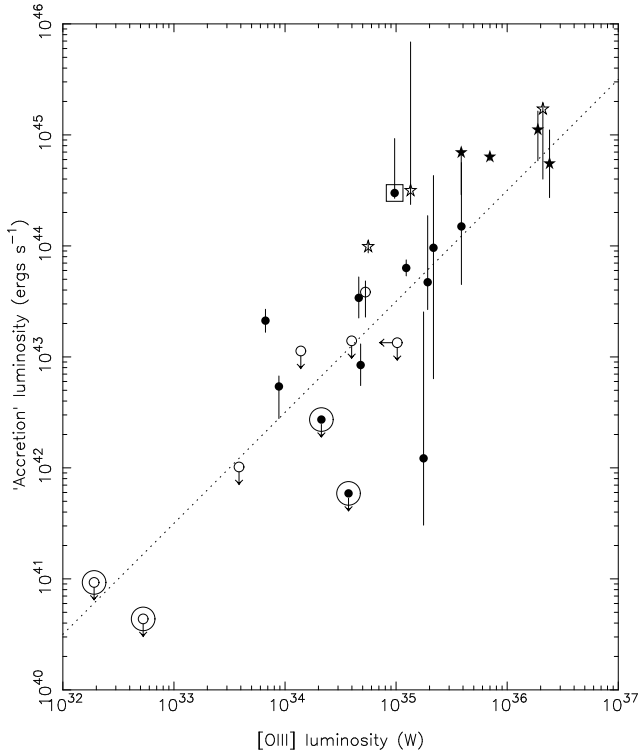}
\caption{X-ray luminosity of the accretion-related component for the
  combined $z<0.5$ sample as a function of power in the [OIII] nuclear
  emission lines for sources with known [OIII] luminosity. Symbols as
  in Fig.\ \ref{radio-xray}. The dotted line is a line of slope unity
  with normalization chosen to pass through the NLRG: it is not in any
  sense a fit to the data. A
  box marks 3C\,295, whose status as a NLRG is in doubt.}
\label{oiii}
\end{figure}

To distinguish between these models we would need independent measures
of the accretion or jet luminosity. One quantity of possible value is
the emission-line luminosity. Rawlings \& Saunders (1991) have argued
that the luminosity in the [OIII] line in FRIIs is well correlated
with jet power, and, if it results from photoionization by the nuclear
continuum, it should correlate well with accretion power as well.
However, Baum, Zirbel \& O'Dea (1995) have shown that the slope and
normalization of radio luminosity-emission line correlations are
different for FRI and FRII populations, and argue that it is possible
that the (mostly low-excitation) FRIs produce emission lines by a
different mechanism, which may not be related to the active nucleus at
all. We cannot rule out the possibility that this is true for the LERG
FRIIs as well (the presence of a LERG population among the FRIIs was
not considered by either Rawlings \& Saunders or Baum \etal) and in
fact Chiaberge \etal\ (2002) show that the FRIs and LERG FRIIs in
their sample lie in a regime where it is possible that the
[OIII]-emitting regions are photoionized not by an AGN but by the
optical (synchrotron) emission from the jet. However,
[OIII] luminosities are readily available for FRIIs and a few
FRIs\footnote{We use the online table collated by Willott \etal\
(1999) and previously discussed in Section \ref{class}, corrected
for the cosmology used in this paper. Given the very large scatter in
line ratios observed in the Willott \etal\ compilation, and the fact
that LERG and BLRG are expected to differ in [OII]/[OIII] line fluxes,
we do not attempt to supplement the true [OIII] fluxes with values
estimated from the luminosities in other lines such as [OII].} and we
plot accretion-related luminosity against [OIII] luminosity for the
objects with available emission-line data in Fig.\ \ref{oiii}.
Although there is clearly significant scatter in this plot (including
error bars, the total range in X-ray luminosities for an [OIII]
luminosity of $10^{35}$ W is up to four orders of magnitude) what is
interesting about it is that the LERG and NLRG (and the few FRIs with
available data) do appear to lie in positions consistent with a linear
relationship between emission-line luminosity and accretion
luminosity. Thus we cannot rule out a model in which there are
conventional active nuclei, with somewhat lower $L_{\rm acc}$ than
would be predicted from their radio luminosity, present in the LERGs
and producing both X-ray and optical line emission: and if we were to
assume that the Rawlings \& Saunders (1991) result holds for LERGs,
this result would favour the model in which it is purely the
relationship between $L_{\rm jet}$ and 178-MHz radio luminosity that
is different about the objects classed as LERGs. This model is also
favoured by the existence of up to two LERGs (3C\,123 and possibly
3C\,295: see Section \ref{class}) that do show detected heavily
absorbed components, albeit as a somewhat low level for their radio
luminosity -- bearing in mind that neither of these is entirely
securely classed as a LERG.

However, it is important to remember that there is no direct evidence
for the existence of an obscuring `torus' in FRII LERGs in general,
just as there is no such evidence in all but a few FRIs. If there is
no torus, then the upper limits on any accretion-related luminosity in
the LERGs would be much lower, and they would no longer lie anywhere
near the straight-line relationship of Fig.\ \ref{oiii}. Thus it is
worth considering the consequences of the alternative model, in which
the objects that lie below the $L_{\rm 178}$-$L_{\rm acc}$
relationship of Fig.\ \ref{lum-lum}, both LERGs and FRIs, do so as a
result of a different accretion mode.

The idea that FRIs may accrete in an advection-dominated mode, leading
to low luminosities from the accretion flow (e.g. Fabian \& Rees
1995), has a long history (e.g. Baum, Zirbel \& O'Dea 1995; Reynolds
\etal\ 1996; Ghisellini \& Celotti 2001; Donato \etal\ 2004, E06,
model 3). The main difficulty from the point of view of radio-galaxy
physics with such a model is that the FRI/FRII dichotomy, and its
dependence on host galaxy properties (Ledlow \& Owen 1996), can be
explained purely in terms of jet power and the interaction with the
environment (e.g. Bicknell 1995). Indeed, an environmental, rather
than nuclear, origin for the FRI/FRII transition seems inescapable
given the existence of (rare) sources with hybrid morphologies
(Gopal-Krishna \& Wiita 2000). Moreover, it is well known that the
parsec-scale jets of FRIs and FRIIs are observationally similar,
arguing against a difference between the classes of source on this
scale (e.g. Pearson 1996; Giovanninni \etal\ 2001). Explaining how a
different accretion mode {\it must} produce a different kpc-scale jet
was thus both difficult and, in some sense, unnecessary. If we accept that
both FRI and FRII kpc-scale structures can be produced by nuclei with
low accretion luminosity (similar to the picture presented by
Chiaberge \etal\ 2002 and Marchesini, Celotti \& Ferrarese 2004) then this problem is removed.
Instead, we can accomodate both FRIIs with low-luminosity accretion
flows, like the LERGs, and FRIs with relatively high-luminosity
accretion flows (Cen A being one example, and the FRI quasar of
Blundell \& Rawlings 2001 another) in a scheme in which environment,
$L_{\rm jet}$ and $L_{\rm acc}$, the last two presumably governed by
parameters of the black hole system like $\dot m$, $M$ and spin, all
play a role.

If the population of LERGs and FRIIs with low accretion luminosity are
identical, how does this fit with the other known distinctive
properties of LERGs? Clearly the low-excitation optical nuclei are
easily explained: there is no luminous AGN to provide the ionizing
continuum necessary for strong optical emission lines (cf. Chiaberge
\etal\ 2002). The situation
with respect to their distinctive environments (Section 1.1) is less
clear. Of the 7 FRII sources we class as LERGs, the {\it Chandra} data
(and earlier studies, in several cases) show 5 (3C\,28, 3C\,123,
3C\,388, 3C\,401, 3C\,438) to lie in rich clusters, consistent with
the findings of Hardcastle (2004), which were based on the galaxy
counts of Harvanek \etal\ (2001). With the exception of 3C\,295, the
most luminous NLRG in the sample (whose status as a NLRG is in doubt:
see Section \ref{class}), none of the NLRG FRIIs lies in a gas-rich
environment. Thus, it remains possible (and this is true whatever the
interpretation of the low accretion luminosities seen in X-ray) that
some of the LERGs have intrinsically low-luminosity jets and an
artifically high low-frequency radio luminosity as a result of their
rich environments and consequent high internal energy densities (cf.
Barthel \& Arnaud 1996), with other morphological peculiarities
(bright jets, weak hotspots, distorted lobes) found in several of
these sources being, similarly, a result of the interaction between
the jet and the rich cluster environment. It is noteworthy that the
objects classified as LERGs that are certainly not in rich
environments (3C\,173.1 and 3C\,200) are entirely morphologically
normal FRIIs. It therefore seems plausible that the cluster-centre
LERGs form a distinct population whose peculiarities are a result of
their environment, and do not need to be explained as a direct result
of their nuclear properties. We note that the cluster-centre FRIIs
would, in general, be expected to be hosted by especially massive
galaxies, with correspondingly large central black holes, and
therefore would have a lower value of the fractional accretion rate
parameter ($\dot m/M_{\rm BH}$: expected to control the mode of
accretion) for a given $\dot m$ than NLRG FRIIs in poorer
environments. Such cluster-centre objects also have a ready supply of
fuel, via Bondi accretion from the dense cluster gas, which is not
available to NLRG in poorer environments.

\subsection{Implications for unified models of FRIIs}
\label{ir}

Our results on LERGs reinforce the idea that these objects cannot be
unified with BLRG and quasars. A prediction to be made from this is
that the objects that {\it are} FRII LERGs seen at a small angle to
the line of sight (these might include BL Lac objects with FRII radio
structures) might share some of the peculiar features of the LERG
population, such as dense environments. This is in principle testable
with X-ray observations of suitable BL Lac objects.

Our results also allow us to comment on the unification of NLRG FRIIs
and BLRG/quasars. The universal detection of a heavily absorbed
nuclear component in the NLRGs is of course {\it qualitatively}
consistent with the expectation from unified models. Like Belsole
\etal\ (2006) we find that the unabsorbed accretion luminosities of
many of the NLRG are similar to the luminosities of BLRG and quasars
in the same range of radio power: this represents a measure of
quantitative consistency with unified model predictions. We see (e.g.
Fig.\ \ref{lum-lum}) that the luminosities of quasars and BLRG tend to
lie slightly above those of the NLRG; Belsole \etal\ attribute the
same tendency in their sample to the presence of beamed, jet-related
X-ray emission in the BLRG and quasars which has been separated out by
differing absorption columns (and would be weaker in any case) in the
NLRG. Although the core/soft X-ray correlation (Fig.\
\ref{radio-xray}) is not well enough constrained to make a definite
statement, estimated corrections to the BLRG/quasar luminosity from
subtracting the jet-related component range from 5 to 40 per cent,
enough to move many of the broad-line objects significantly lower on
Fig.\ \ref{lum-lum}. The photon indices of the absorbed components in
radio galaxies, where well constrained, occupy the same range as those
of the unabsorbed power laws fitted to broad-line objects, which is
also consistent with the simple unified model expectation.

\begin{figure}
\epsfxsize 8.4cm
\epsfbox{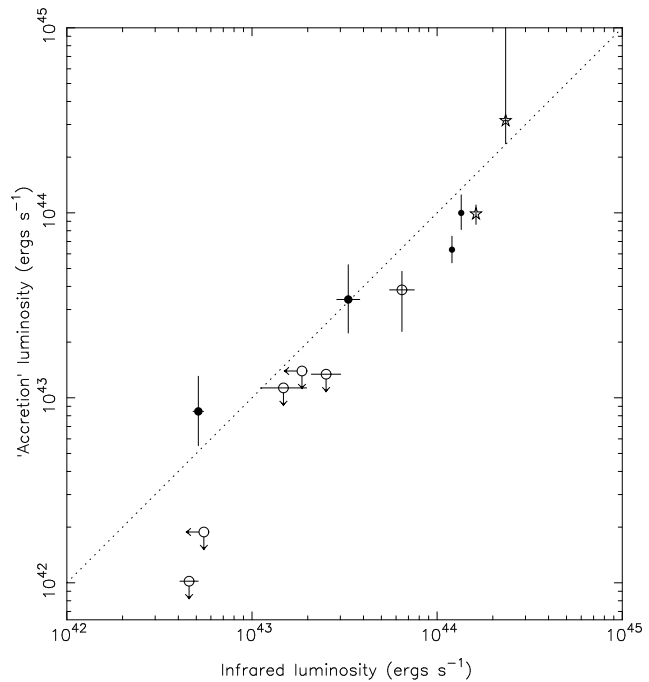}
\caption{X-ray luminosity of the accretion-related component for the
  combined $z<0.5$ sample as a function of luminosity in the
  mid-infrared ($\nu L_\nu$ at 15 $\mu$m), taken from Ogle \etal\ (2006), for 12 sources with both
  infrared and X-ray measurements or upper limits. The dotted line is
  the line of {\it equality} between the two luminosities, and is {\it
  independent} of the data.}
\label{ogle}
\end{figure}

A few of the detected NLRG FRIIs in our sample have accretion-related X-ray
luminosities $<10^{43}$ ergs s$^{-1}$, significantly below the $\sim
10^{44}$ ergs s$^{-1}$ of the least luminous detected BLRG. This is
reminiscent of the results of Ogle, Whysong \& Antonucci (2006) who
have recently used {\it Spitzer} data to show that some radio galaxies
do not have a luminous ($>10^{44}$ ergs s$^{-1}$) mid-infrared
nucleus. Some of the `mid-infrared weak' sources of Ogle \etal\ are
objects that we class as LERGs, which we might expect to form a
separate population, but others are NLRGs, including two in our
sample, 3C\,192 and 3C\,436. Our X-ray data for these sources allow us
to say with certainty that, while they may not host a `luminous
accretion disc' by the definition used by Ogle \etal, they certainly
do host a heavily obscured nuclear component and would appear quite
different in the X-ray if the obscuration were removed. Occam's razor
suggests that these NLRG would be expected to have BLRG counterparts
of similar accretion luminosity. Having said this, the correlation
between the accretion and mid-IR luminosities for the few sources in
common between the sample of Ogle \etal\ and ours (Fig.\ \ref{ogle})
is remarkable: the sources that they find to have low infrared
luminosity have exactly corresponding low X-ray luminosity. This
result provides a foretaste of what may be expected when comparisons
of X-ray and infrared data for larger numbers of sources are
available, as discussed by Belsole \etal 

Finally, we note that the fact that the LERGs in common between the
two samples lie near or below the line of equality in Fig.\ \ref{ogle}
strongly suggests that we have not grossly underestimated the true
LERG accretion luminosity, at least in these sources. If the column
densities for these objects were $\ga 10^{24}$ cm$^{-1}$, the upper
limits on the `hidden' nuclear luminosities would be 1 or more orders
of magnitude higher than the upper limits we have plotted, and if the
true luminosities were close to the upper limits, we should see a
correspondingly large nuclear luminosity re-radiated in the infrared,
which is not observed. This retrospectively justifies our choice of a
limiting column density of $10^{23}$ cm$^{-1}$.

\section{Summary and conclusions}

Our results and conclusions can be summarized as follows.

\begin{itemize}
\item While normal NLRG universally show a heavily absorbed component
  of nuclear X-ray emission, this is rarely found in low-excitation
  FRIIs (LERGs): thus the LERGs are observationally similar in their
  properties to the FRIs studied by E06, mostly showing only
  unobscured X-ray emission that we attribute to the parsec-scale jet.

\item Discounting the possibility of absorbing columns significantly
  greater than $10^{23}$ cm$^{-2}$, we can place upper limits on the
  luminosity of any undetected accretion-related component present in
  the LERGs which are substantially below the values for detected
  NLRG of similar total radio luminosities.

\item Some lines of evidence, notably a reasonable correlation between
  accretion luminosity and [OIII] emission line luminosity that would
  exist if the accretion luminosities are close to the upper limit
  values, support the idea that (at least some of) the LERGs do have
  absorbed nuclear components from a conventional accretion disc, but
  at a level lower than would be inferred from their radio
  luminosities. This would imply that the LERGs are of the order of 10
  times more efficient than the NLRG at producing low-frequency radio
  emission for a given accretion luminosity, which may relate to the
  rich environments of many LERG sources.

\item However, there remains a strong possibility that some, or most,
  FRII LERGs accrete in a radiatively inefficient mode, and have
  little or no X-ray accretion-related emission, as has been
  previously argued for FRI sources. Allowing both FRIs and FRIIs to
  be generated by an active nucleus powered by a radiatively
  inefficient accretion flow removes some of the problems of earlier
  models along these lines. The FRI/FRII dichotomy in this picture
  would remain a function of jet power and source environment, while
  the low-excitation/high-excitation dichotomy would be controlled by
  the accretion mode and thus presumably by black hole mass and
  accretion rate. There is no requirement for dusty tori in the
  LERGs, but we cannot rule out the possibility that they exist.

\item Finally, we show that the accretion luminosities of the NLRG and
  broad-line objects (BLRG and quasars) in our sample are roughly
  consistent. Some NLRG have low-luminosity absorbed nuclei, but we
  argue that such objects must still participate in normal unified models.

\end{itemize}

\section*{Acknowledgements}

MJH thanks the Royal Society for a research fellowship. This work is
partly based on observations obtained with {\it XMM-Newton}, an ESA
science mission with instruments and contributions directly funded by
ESA Member States and the USA (NASA). We are grateful to Marco
Chiaberge and to the anonymous referee for comments that helped us to
improve the paper.

\appendix
\section{Notes on individual sources}

\subsection{3C\,28}

The core of this low-excitation source is undetected in both radio and
X-ray. We derive X-ray upper limits from the local count density, which is
high as the source lies in the centre of one component of the merging
cluster Abell 115 (e.g. Gutierrez \& Krawczynski 2005). Limits on
luminosities in absorbed and unabsorbed components were derived assuming
photon indices of 1.7 and 2.0 respectively.

\subsection{3C\,47}

This quasar is reasonably well fitted ($\chi^2 = 62/48$) with a single
power law model with $\Gamma = 1.5 \pm 0.1$, but when a second power
law is added the fit is improved and the 90 per cent error on the
normalization is non-zero, so we adopt that model. The
single-power-law fit is consistent with earlier results (Lawson \etal\
1992).

\subsection{3C\,79}

To our knowledge this is the first X-ray spectrum to be measured for
this source. An unabsorbed and absorbed component are required by the data.

\subsection{3C\,109}

The moderate intrinsic absorption we find in the spectrum of this BLRG
is in good agreement with the results of earlier {\it ROSAT} and {\it
ASCA} observations (Allen \& Fabian 1992; Allen \etal\ 1997) given the
slightly different Galactic absorption column we adopt. The photon
indices and spectral normalizations are also consistent within the
joint 90 per cent confidence limits. We find no evidence for the iron
line reported by Allen \etal\ (1997), although the signal to noise is
relatively low in the {\it Chandra} data because of the annular
extraction region we use.

\subsection{3C\,123}
\label{123}

The Galactic column density used for this source is taken from the
earlier analysis of the {\it Chandra} data by
Hardcastle \etal\ (2001) and the dominant (absorbed) power law model
has similar parameters to the one fitted there. The large amount of
Galactic reddening (the source is known to lie behind a molecular
cloud) makes its classification as a low-excitation object uncertain.
Chiaberge \etal\ (2002) show that its position on their diagnostic
plane (equivalent width of [OIII] vs. optical to radio core flux
ratio) is more consistent with a classification as a NLRG, and this
would be consistent with the observed luminous, heavily absorbed
nuclear X-ray component.

\subsection{3C\,173.1}

The small number of counts in the core of this low-excitation object
means that only the normalization of X-ray models is constrained.
Accordingly, the upper limit on the normalization of an obscured model
is derived assuming that such a model produces {\it all} the observed counts.

\subsection{3C\,192}

When this source was fitted with the standard two-component power-law
model there were prominent residuals in the range 0.5-1.0 keV,
suggesting the addition of a thermal component. We obtained good fits
by adding a MEKAL model with fixed $kT = 0.5$ keV and 0.5 solar abundance.
To our knowledge the X-ray spectrum of this source has not been
discussed elsewhere.

\subsection{3C\,200}

This source was classed as a NLRG by Belsole \etal\ (2006),
following LRL, but the high-quality spectra of Laing
\etal\ (1994) show it to be a LERG, and we adopt that classification
here. Our spectral fits are, unsurprisingly, consistent with those of
the analysis of the same data by Belsole \etal

\subsection{3C\,215}

The photon index for the single-power-law fit to this quasar is
consistent with that found with {\it ASCA} by Reeves \& Turner (2000),
but its luminosity in our measurements is nearly a factor 2 lower,
probably indicating variability.

\subsection{3C\,219}

We see no evidence in the spectrum of this broad-line radio galaxy for
the excess absorption fitted by Sambruna \etal\ (1999) to {\it ASCA}
data for the source, but there is a good deal of spatially resolved
X-ray structure seen in the {\it Chandra} data (Comastri \etal\ 2003)
which may have confused their fits.

\subsection{3C\,223}

The {\it XMM-Newton} data we use here were analysed by Croston et al.
(2004). They fitted a model consisting of an unabsorbed power law, plus an
absorbed power law and iron line; however, they fixed the photon index of
the second power-law at $\Gamma = 1.5$, which leads to differences in the
best-fitting spectral parameters. An apparent discrepancy in the flux
densities obtained by the two analyses can be attributed to changes to the
{\it XMM} calibration files since the work of Croston et al.

\subsection{3C\,249.1}

Both {\it Chandra} and {\it XMM-Newton} data exist for this quasar: we
use only the {\it XMM} data, since there are more counts and no pileup
issues to contend with. The best-fitting model includes a second power
law with moderate intrinsic absorption, and a narrow Gaussian with rest-frame
energy 6.42$_{-0.06}^{+0.03}$ keV. The parameters of the
double-power-law model are very similar to those fitted by Piconcelli
\etal\ (2005) to the same X-ray data.

\subsection{3C\,284}

The {\it XMM-Newton} data we use here were analysed by Croston et al.
(2004). Their best-fitting model parameters are in reasonable agreement
with ours, although they fixed the photon index at $\Gamma = 1.5$, rather
than 1.7 as used in this work. Apparent discrepancies in the absorption
column affecting the second power-law component and in the flux densities
result from changes in the {\it XMM} calibration files, as for 3C\,223.

\subsection{3C\,285}

Ours is the first X-ray spectrum to be measured for the nucleus of
this source. A two-component model with absorbed and unabsorbed power
laws and a strong Gaussian component with rest-frame energy
$6.42\pm0.05$ keV are required, although the data are not good enough
to constrain the photon indices of the power-law components.

\subsection{3C\,295}

The longer observation we use allows us to fit a more sophisticated
model to the data than was possible for Harris \etal\ (2000): the
two-component best-fitting model explains the inverted spectrum found
by Harris \etal\ when fitting a single power law with Galactic absorption.
The addition of a narrow Gaussian component with rest-frame energy
$6.4^{+0.1}_{-0.5}$ keV gives a significant improvement to the fits.

\subsection{3C\,303}

The power-law index we fit to this BLRG is consistent with the results
from two sets of archival {\it ASCA} data reported by Kataoka \etal\
(2003). The luminosity inferred from the flux reported by Kataoka
\etal\ is a factor $\sim 1.5$ higher than we observe, which may imply
variability.

\subsection{3C\,346}

3C\,346 is the only source classed as an FRI in our new sample.
Worrall \& Birkinshaw (2005) fitted a model with a small amount of
intrinsic absorption to the {\it Chandra} data we use, and accordingly
found a slightly steeper photon index and higher unabsorbed 1-keV flux
density.

\subsection{3C\,351}

The parameters of the best-fitting model are similar to those found
in the analysis of the same data in
Hardcastle \etal\ (2002), but the photon index of the soft power law
is even steeper than found in that paper. The change is presumably a
result of updated responses for the ACIS-S detector.

\subsection{3C\,388}

The fits to this source are consistent with those in E06.

\subsection{3C\,401}

The parameters we fit for this source are consistent within the errors
with those found by Reynolds, Brenneman \& Stocke (2005) in their
analysis of the same {\it Chandra} data.

\subsection{3C\,436}

The X-ray spectrum of this NLRG has not previously been
investigated. In addition to the standard two-power-law model,
residuals around 5 keV are reduced with the addition of a narrow
Gaussian with rest-frame energy $6.4_{-0.1}^{+0.2}$ keV.

\subsection{3C\,438}

Our luminosity and photon index for this radio galaxy are consistent
with the analysis of the same data in Donato \etal\ (2004).

\end{document}